\documentclass[12pt,a4paper]{article}
\usepackage{amssymb}
\usepackage[T2A]{fontenc}
\usepackage[cp866]{inputenc}
\usepackage[russian,english]{babel}
\usepackage[unicode]{hyperref}
\begin{document}
\def\emline#1#2#3#4#5#6{%
       \put(#1,#2){\special{em:moveto}}%
       \put(#4,#5){\special{em:lineto}}}
\def\newpic#1{}
\voffset=-2cm
\begin{center}
{\Large The pedagogical value of the four-dimensional picture I:
Relativistic mechanics of point particles}\\[3mm]
{B. P. Kosyakov} \\[3mm] 
{Russian Federal Nuclear Center--VNIIEF and Sarov Institute of Physics {\&} Technology, Sarov, 607190 Nizhnii Novgorod 
Region, Russia\\
} 
\end{center}
\begin{abstract}
\noindent
{\small 
We outline two subjects of relativistic mechanics: (i) the set of allowable world 
lines, and (ii) the origin of  
the relativistic law of dynamics governing point particles. 
We show that: (i) allowable world 
lines in the classical theory of particles and fields are quite simple geometric objects as opposed to their associated three-dimensional trajectories, and (ii) Newton's second law requires neither modification nor generalization, it should be only smoothly embedded in the four-dimensional geometry of Minkowski spacetime to yield the dynamical law for relativistic particles.}

\end{abstract}

\section{Introduction}
\label
{Introduction}
Hermann Minkowski in his talk at the 80th Assembly of German Natural Scientists 
and Physicians on 21 September 1908 asserted:  
``Space by itself, and time by itself, are doomed to fade away into mere shadows, 
and only a kind of union of the two will preserve an independent reality''~\cite{Minkowski}.    
He anticipated that physicists will be inspired with the idea of 
four-dimensional spacetime, and their way of thinking will be rearranged in accordance with this discovery.
But the anticipation comes true not very soon.   
Einstein dispensed with the four-dimensional spacetime up to 1913 because he regarded
this geometric idea as `superfluous learnedness' (\"uberfl\"ussige Gelehrsamkeit)~\cite{Abraham}.
He began to consider seriously the four-dimensional picture when he realized that 
{\it gravity} is a manifestation of {\it  spacetime warping}~\cite{Einstein13}. 

The experimental confirmation of effects predicted by general relativity resulted in the consensus that the four-dimensional picture is an integral part of the theory of gravitation.
Then the covariant perturbation techniques of quantum electrodynamics, developed  in the late 1940s, offered a means for making  such calculations which would be far beyond the human possibilities without resort to this 
machinery.
Precise measurements showed that the agreement between the measured and the calculated quantities is within an accuracy of $10^{-10}$.
Four-dimensional quantum field theory began rapidly develop culminating in the proof of the theorems on  $PCT$-invariance and connection between spin and statistics.
The leading part of the physics community realized that the concept of spacetime represents a profound paradigm for description of our world.  

How did this realization reflect in teaching of theoretical physics?
General relativity is  from the outset covered in the four-dimensional language.
In contrast, electrodynamics, garnished with some elements of special relativity, is reproduced in the physics curriculum almost unchanged from generation to generation in the tradition of a century ago.

In an attempt to bridge the gap between the conceptual and mathematical levels of these disciplines we begin on this series of papers,
lumped together as ``The pedagogical value of the four-dimensional picture'', bearing in mind that  electrodynamics shapes the worldview and way of thinking of every physicist. 

It is the author's opinion that the rudimentary foundation of the 4D culture should be laid still in secondary school.
Undergraduate students in physics should have a clear idea of 
the light cone and elementary concepts of pseudoeuclidean space (four-vectors, 
straight lines, hyperplanes, scalar product, etc), and confidently use this geometric tools. 
Should this be the case,  the physics curriculum  would be amenable to a
truly four-dimensional treatment. 

In the present paper, we outline two topics in relativistic mechanics which are not fully, if at all, elucidated in textbooks: (i) the set of allowable world lines in the classical theory of particles and fields, and (ii) the origin of  the 
relativistic dynamical law governing point particles and the afforded by this law equation of motion for Galilean particles.

\section{World lines and spatial trajectories}
\label
{worldlines}
The history of a particle in space-time  is depicted by a curve 
extending from the remote past to the far future.
This curve is called the {\it world line}.

A convenient way to define a curve in three-dimensional Euclidean space ${\mathbb E}_3$ is to consider a smooth {\it mapping} ${\bf z}:\,\, t\mapsto {\bf z}(t)$ of an interval $[t_1,t_2]$ on the real axis ${\mathbb{R}}$ into ${\mathbb E}_3$.
Such mappings are called {\it parametrized curves}.
The concept of world lines is defined similarly.
By the term `world line' is meant 
a vector-valued smooth function $ z^\mu (\tau) $ whose argument 
is understood as an evolution parameter ranging from  $ - \infty $ to $ + \infty $.
 
The parametrization of the world line can be chosen so as to alleviate the 
problem as much as possible.
For example, we may identify the parameter $ \tau $ with laboratory time $ t $ 
of a particular inertial frame.
Then $ z^\mu (t) $ is decomposed into $ z^\mu (t)= \left(t,{\bf z}(t)\right) $, 
where the spatial component, denoted by a boldface character, is a parameterized
curve $ {\bf z} (t) $  called the {\it trajectory} of the particle.
Locally, the trajectory is characterized by the velocity $ {\bf v} $ and the acceleration
$ {\bf a} $,
\begin{equation}
{\bf v}=\frac{d{\bf z}}{dt}\,\,,
\quad
{\bf a}=\frac{d{\bf v}}{dt}\,\,.
\label
{velocity,acceleration-df}
\end{equation}                          

Relativistic problems are usually formulated by taking an invariant variable as 
the parameter of evolution.
It is convenient to parametrize world lines of a particle by its {\it proper time}.
If the particle is equipped with the clock identical to that of a stationary 
Lorentz observer, then the proper time $ s $ of the particle is defined as 
the time read from this moving clock. 
Consider an inertial frame in which the particle is at rest at a given instant.
This frame of reference is called the {\it rest frame} or {\it instantaneously 
comoving} frame.
In this system, the line element $ dz^2 $ 
equals\footnote{We use the system of units in which the speed of light is equal to 1.
With these units, length and time are measured in the same units, say, in seconds, 
the velocity  ${\bf v}$ is dimensionless, and the acceleration ${\bf a}$ has  
dimension inverse of time dimension.
We adopt the metric $\eta_{\mu\nu}$ with the signature $(+1,-1,-1,-1)$.}  the squared local proper time interval $ ds^2 $.
Since $ dz^2 $ is an invariant quantity, it can be expressed in terms of
the laboratory frame coordinates: $ dz^2 = dt^2-d {\bf z}^2 $,
which implies
\begin{equation}
ds=\sqrt{dt^2-d{\bf z}^2}=dt\,\sqrt{1-{\bf v}^2}\,\,.
\label
{ds=dt-sqrt(1-v-2)}
\end{equation}                          
With the use of the  {\it Lorentz factor}
\begin{equation}
\gamma=\frac{1}{\sqrt{1-{\bf v}^2}}\,\,,
\label
{gamma=Lorentz-factor}
\end{equation}                          
(\ref{ds=dt-sqrt(1-v-2)}) becomes
\begin{equation}
dt=\gamma\,ds\,\,.
\label
{dt=gamma-ds}
\end{equation}                          

Locally, the world line, parametrized by the proper time $s$, is characterized by the four-velocity $ v^\mu $ and
the four-acceleration $ a^\mu $, 
\begin{equation}
{v}^\mu=\frac{d{z}^\mu}{ds}\,\,,\quad
{a}^\mu=\frac{d{v}^\mu}{ds}\,\,.
\label
{four-velocity-df} 
\end{equation}                          

Since $dz^2=ds^2$, we deduce from the first formula of (\ref{four-velocity-df}) the identity  
\begin{equation}
{v}^2=1\,\,.
\label
{four-velocity-2=1}
\end{equation}                          
The four-velocity 
$v^\mu$ is by definition a {\it tangent}  vector.
Differentiating (\ref{four-velocity-2=1}) with respect to $ s $, we find
\begin{equation}
{v}\cdot a=0\,\,.
\label
{four-velocity-times-four-acceleration=0}
\end{equation}                          
Thus, the four-acceleration $ a ^ \mu $ is always perpendicular to the world line.

Using (\ref{ds=dt-sqrt(1-v-2)}) and (\ref{gamma=Lorentz-factor}), it is easy to
show that, in a particular inertial frame of reference,
\begin{equation}
{v}^\mu=\gamma\left(1,\,{\bf v}\right),
\label
{four-velocity-in terms-velocity}
\end{equation}                          
\begin{equation}
a^\mu=\gamma^2\left(\left({\bf a}\cdot{\bf v}\right)\gamma^2,\,{\bf a}
+
{\bf v}\left({\bf a}\cdot{\bf v}\right)\gamma^2\right).
\label
{four-acceleration-in terms-velocity}
\end{equation}                          

When considering relativistic kinematics, one usually assumes that particles move
at speeds less than the speed of light, in other words, the world lines are 
described by timelike curves.
A curve is {\it timelike} if the tangent vector is timelike at any point 
of this curve.
A curve is {lightlike}/{spacelike} if the tangent 
vector is lightlike/spacelike  at any point of it.

What kind of world lines is allowable in classical physics?
Many relativistic expressions, such as (\ref{four-velocity-in terms-velocity}) 
and (\ref{four-acceleration-in terms-velocity}), contain the Lorentz factor 
(\ref {gamma=Lorentz-factor}) which becomes $ \infty $ as $ | {\bf v} | \to 1 $.
Infinite value of $ \gamma $ signals that a particle 
cannot reach velocities greater than the velocity of light.
Geometrically, a faster-than-light particle moves along a spacelike world line.
Therefore, any curve with smoothly joined
timelike and spacelike fragments must be excluded from the set of allowable world lines.

A similar restriction 
imposes the classical {\it causality} argument.
If faster-than-light particles would exist, then it would be possible for an
observer to send signals to his own past.
A sequence of events involving such an impact on the past is
called a {\it causal cycle}.
The admission of superluminal signals would result in a patently absurd process
in which an observer could cause his own destruction in the past which, in turn, 
prevents the destructive signal from being sent.
Causal cycles are depicted as closed world lines.
If a cycle is smooth, then it necessarily contains spacelike fragments.
This consideration seems to forbid spacelike 
fragments from world lines in the classical relativistic theory.

Is it possible to weaken the {\it smoothness} requirement? 
Is a jump of the tangent vector at some point of the world line admissible?
To answer this question, let us take a closer look at such piecewise smooth world lines.

\begin{figure}[htb]
\begin{center}
\begin{minipage}[t]{154mm}
\unitlength=1.00mm
\special{em:linewidth 0.4pt}
\linethickness{0.4pt}
\begin{picture}(111.00,45.00)
\put(10.00,10.00){\vector(0,1){7.00}}
\bezier{52}(10.00,17.00)(10.00,25.00)(12.00,30.00)
\bezier{60}(12.00,30.00)(14.00,34.00)(14.00,45.00)
\put(30.00,40.00){\vector(0,-1){7.00}}
\bezier{44}(30.00,33.00)(30.00,28.00)(32.00,22.00)
\bezier{48}(32.00,22.00)(35.00,13.00)(35.00,10.00)
\put(10.00,6.00){\makebox(0,0)[cc]{particle}}
\put(35.00,6.00){\makebox(0,0)[cc]{antiparticle}}
\put(80.00,10.00){\vector(0,1){6.00}}
\put(97.00,16.00){\vector(0,-1){6.00}}
\bezier{40}(80.00,16.00)(80.00,22.00)(82.00,26.00)
\bezier{52}(88.00,38.00)(84.00,30.00)(82.00,26.00)
\put(88.00,38.00){\makebox(0,0)[cc]{$\bullet$}}
\put(88.00,43.00){\makebox(0,0)[cc]{$\rm A$}}
\put(112.00,36.00){\vector(0,1){7.00}}
\put(131.00,43.00){\vector(0,-1){7.00}}
\bezier{56}(131.00,36.00)(131.00,29.00)(129.00,22.00)
\bezier{56}(129.00,22.00)(126.00,13.00)(123.00,10.00)
\bezier{76}(123.00,10.00)(115.00,22.00)(114.00,26.00)
\bezier{40}(114.00,26.00)(112.00,32.00)(112.00,36.00)
\put(123.00,10.00){\makebox(0,0)[cc]{$\bullet$}}
\put(123.00,6.00){\makebox(0,0)[cc]{$\rm B$}}
\bezier{56}(88.00,38.00)(89.00,33.00)(95.00,24.00)
\bezier{32}(95.00,24.00)(97.00,20.00)(97.00,16.00)
\end{picture}
\caption{World lines of particles and antiparticles. 
$\Lambda$-shaped and $V$-shaped curves
}
\label{crann}
\end{minipage}
\end{center}
\end{figure}
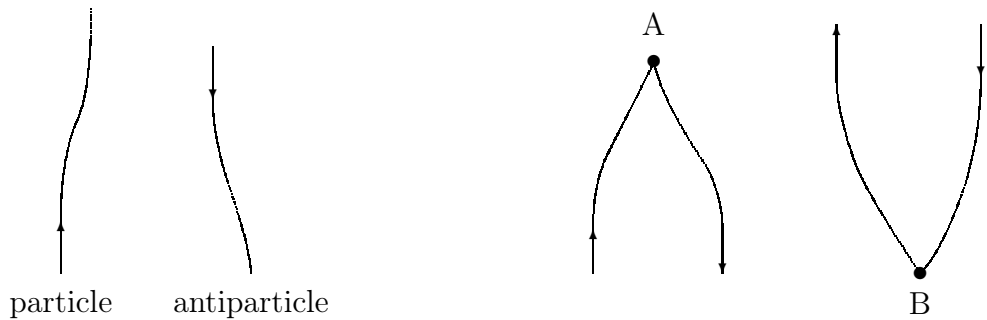

Given a particle moving along a timelike world line oriented from the past to the future, its antiparticle may be thought of as an object identical to it in every respect but moving back in time  \cite{Feynman}.
That is, the antiparticle's world line is oriented from the future to 
the past (left plot of Figure \ref{crann}).
The annihilation of a pair that occurs at a point ${\rm A}$ 
is depicted as a $\Lambda$-shaped world line of a {\it single} particle that 
runs initially from the remote past to the future 
up to the point ${\rm A}$ and then returns to the remote past.
Likewise, the birth of a pair occurring at a point ${\rm B}$ is 
given by a $V$-shaped world line of a {single particle} 
that runs initially from the far future to the 
past up to the point ${\rm B}$ and then returns to the far future, 
as in Figure \ref{crann} right.
Creations and annihilations of  pairs of particles and their antiparticles, which are possible in the quantum picture,  
are strongly banned in the classical picture.
Thus, the mathematical condition for precluding the occurrence of $V$-shaped and $\Lambda$-shaped world lines is equivalent to the physical requirement to forbid creation and annihilation processes.

Causal cycles can be built from $V$-shaped and $\Lambda$-shaped timelike curves.
Therefore, such curves are inappropriate to the classical theory.
They are automatically eliminated from consideration if we accept the condition that all solutions to the dynamical equations belong to a 
set of smooth timelike curves, that is, the set of allowable world lines is deprived from sectionally smooth curves.

In nature, there are particles which travel at luminal velocities, even though they were never accelerated from a slower speed.
These particles begin moving with the speed of light immediately upon creation and remain in this state.
World lines of these particles are depicted by null curves. 
Such objects presents no problems with causality.
Lightlike smooth curves may also be included in the set of allowable world lines.

We thus see that allowable {world lines} of classical particles are rather {trivial topological} concepts.  
If we pull a timelike world line  on the ends, we get a straight line
directed along the time axis in some inertial frame.
Furthermore,  allowable world
lines are simple from the differential-geometric point of view.
If we assume that all world lines in some classical field theory
are {infinitely differentiable} parameterized curves, this assumption does not deplete the content of this theory.
 
In contrast, the {trajectory} is a {nontrivial} topological concept.
A particle can move along an extremely entangled path in space.
In the language of the theory of knots, this path is a curve with an arbitrary number $k$ of nontrivial interweavings.
If we pull it on its ends, we
obtain $ k $ tied knots, while the same operation with the
corresponding world line results in no knot. 
The reason for this is that a smooth timelike curve never changes its orientation relative to the time axis, and hence it cannot form  loops potentially
suitable for the occurrence of tied knots.
Note also that any curve in  Euclidean space ${\mathbb E}_n$ is free of  knots when  $n\ge 4$  \cite{knots}. 
A further aggravating fact is that the trajectory is a complicated differential-geometric concept, because there is no reason for eliminating  three-dimensional curves  
with cusps.
For example, the trajectory of a stone moving up 
and down along a vertical direction involves a cusp.
This curve results from a sequence of smooth trajectories of the stone thrown at an angle $\alpha$ to 
the horizontal as $\alpha\to\frac{\pi}{2}$.

Thus, the description of the set of allowable trajectories is a cumbersome task.
It does not add novel results to those which are found in studying the set of allowable world lines.

\section{The dynamical law for a relativistic particle}
\label
{General law}
We begin with Newton's second law in its original form,
\begin{equation}
\frac{d{\bf p}}{dt}={\bf f}\,\,.
\label
{newton-orig-p}
\end{equation}                        
Is this equation completely unsuitable for the description of a relativistic particle? 
What is the true law of dynamics in special  relativity?
Is something additional necessary for obtaining this law?
{In this point, many textbooks mislead the 
newcomer.  
For example, Hartle asserts:  
``The laws of Newtonian mechanics have to be changed to be consistent with 
the principles of special relativity'' \cite{Hartle77}, and 
``The objective of relativistic mechanics is to introduce the analog of 
Newton's second  law ${\vec F}=m{\vec a}$. 
There is nothing from which this 
law can be {\it derived}, but plausibly it must satisfy certain properties'' \cite{Hartle85}}.

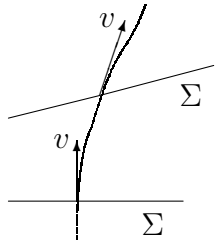
\begin{figure}[htb]
\begin{center}
\unitlength=1mm
\special{em:linewidth 0.4pt}
\linethickness{0.4pt}
\begin{picture}(38.00,34.00)
\bezier{60}(20.00,5.00)(20.00,16.00)(22.00,20.00)
\bezier{44}(22.00,20.00)(24.00,27.00)(26.00,30.00)
\put(20.00,10.00){\vector(0,1){8.00}}
\put(23.00,24.00){\vector(1,3){3.33}}
\emline{23.00}{24.00}{1}{38.00}{28.00}{2}
\emline{23.00}{24.00}{3}{11.00}{21.00}{4}
\emline{11.00}{10.00}{5}{34.00}{10.00}{6}
\put(30.00,7.00){\makebox(0,0)[cc]{$\Sigma$}}
\put(35.00,24.00){\makebox(0,0)[cc]{$\Sigma$}}
\put(18.00,18.00){\makebox(0,0)[cc]{$v$}}
\put(24.00,34.00){\makebox(0,0)[cc]{$v$}}
\bezier{36}(26.00,30.00)(28.00,33.00)(29.00,36.00)
\end{picture}
\caption{The hyperplane $ \Sigma $ perpendicular to the world line}
\label
{hyperplane}
\end{center}
\end{figure}

However, a closer look at the subject shows that the relativistic dynamical law is just Newton's second law  (\ref {newton-orig-p})  {\it embedded} into the 
geometry of four-dimensional spacetime.
We restrict our consideration to the case that a particle moves along a timelike world line.
Note that (\ref{newton-orig-p}) becomes a {\it strict} law as ${\bf v}\rightarrow 0$. 
Granting that an {instantaneously comoving} inertial frame 
is given at some instant, one can precisely predict the evolution of the 
particle in this frame during an ensuing evanescent time interval.
In the geometric language,  equation (\ref{newton-orig-p}) is a
strict vector relation on a hyperplane $\Sigma $ perpendicular to the  world line. 
Meanwhile the hyperplane $\Sigma $ tilts together with its normal $v^{\mu }$ as
one moves along the curve, see Figure  \ref{hyperplane}.
Such hyperplanes constitute a one-parameter family.
For the embedding, we need an operator 
$\stackrel{\scriptstyle v}{\bot }$\thinspace\ that 
projects four-vectors 
on hyperplanes $\Sigma $ perpendicular to the world line. 
The required operator is 
\begin{equation}
\stackrel{\scriptstyle v}{\bot}_{\hskip0.5mm\mu\nu}\,=\,
{\eta}_{\mu\nu}-\frac{v_\mu v_\nu}{v^2}\,\,.
\label
{projector-def}
\end{equation}                        
The form of the projector
(\ref{projector-def}) is the same for any parametrization of the world line, that is,
if
$v_\mu={dz_\mu/ds}$ is replaced by ${\dot z}_\mu={dz_\mu/d\tau}$, this
leaves the form of $\stackrel{\scriptstyle v}{\bot}$\thinspace\ unchanged.

The time axis in the instantaneously comoving frame is parallel to the tangent 
of the world line, and hence $dt$ is equal to $ds$.
Accordingly, in this frame, the derivative with respect 
to $t$ may be replaced by the derivative with respect to $s$. 

Given the Newtonian force  ${\bf f}$ in every hyperplane $\Sigma$, one 
can unambiguously construct a four-dimensional vector $f^{\mu }$. 
With this in mind,  we take the instantaneously comoving frame, and write the decomposition of $f^\mu$ in this frame:  
\begin{equation}
f^\mu=(0,\,{\bf f})\,\,. 
\label
{f-rest-syst}
\end{equation}                        
The components of $f^{\mu }$ in an arbitrary inertial frame can be found 
from (\ref{f-rest-syst}) through the use of an appropriate Lorentz boost.
$f^{\mu }$ is called  the {\it Minkowski force} or {\it four-force}.

In a rest frame, we have the decomposition $v^\mu=(1, {\bf 0})$.
Combined with (\ref{f-rest-syst}), this gives
\begin{equation}
f\cdot v=0\,\,. 
\label
{f-cdot-v}
\end{equation}
Since the scalar product of two four-vectors is Lorentz invariant, we conclude that the four-force $f^{\mu }$ is perpendicular to the four-velocity $v^\mu$ in any frame.

We should finally introduce the notion of {\it four-momentum} $p^{\mu}$.
But our concern here is only with the derivative of $p^{\mu}$ with respect to the 
proper time, $dp^\mu/ds$.
Let us require that, in the 
instantaneously comoving frame, the space components  $dp^i/ds$ be identical to the respective 
components of $d{\bf p}/dt$ in Newton's second law (\ref{newton-orig-p}), while the time component $dp^0/ds$ is left indeterminate.

The embedding of Newton's second law (\ref{newton-orig-p}) in 
hyperplanes perpendicular to the world line is given by
\begin{equation}
\stackrel{\scriptstyle v}{\bot}_{\hskip0.5mm\mu\nu}\!\left(\frac{dp^\nu}{ds}-
f^\nu\right)=0\,\,.
\label
{newton4}
\end{equation}                        
This is the desired {\it dynamical law} for a relativistic particle. 
In symbolic form,   
\begin{equation}
\stackrel{\scriptstyle v}{\bot}({\dot p}-f)=0\,\,.
\label
{newton-symbl}
\end{equation}                        

The presence of  
$\stackrel{\scriptstyle v}{\bot}$ in 
(\ref{newton-symbl}) suggests that we have three independent 
equations.
Indeed, when contracted with $v^\mu$, the four equations 
reveal a 
linear relation resulting from the identity
\begin{equation}
v^\mu\left({\eta}_{\mu\nu}-\frac{v_\mu v_\nu}{v^2}\right)=0\,\,.
\label
{ident-newton-tql}
\end{equation}                        

Newton postulated that  ${\bf p}$ is proportional to ${\bf v}$,
\begin{equation}
{\bf p}=m{\bf v}\,\,.
\label
{p=mv-3}
\end{equation}                        
The coefficient of proportionality $ m $ measures inertia of the particle, 
and is called the {\it Newtonian mass}.
This postulate is not the only possible.
Indeed, (\ref{p=mv-3}) can be modified in various ways by  adding terms proportional to acceleration 
${\bf a}$ and higher derivatives of ${\bf v}$.
However, (\ref{p=mv-3}) offers the advantage over all other relationships between the momentum and kinematical characteristics of the particle in that it is the  simplest Euclidean relationship. 
In fact, (\ref{p=mv-3}) is  the defining equation for a certain class of physical
objects with the momentum of just this form.
We call such objects {\it Galilean particles}, or simply
{\it particles}.
Substitution of (\ref{p=mv-3}) in (\ref{newton-orig-p}) gives the {\it equation
of motion} for a Galilean particle
\begin{equation}
m{\bf a}={\bf f}\,\,.
\label
{newton-e-m}
\end{equation}

In the four-dimensional picture, this notion is slightly reformulated.
We refer to the {Galilean particle} as 
a point object which  possesses 
the four-momentum   
\begin{equation}
p^{\mu}=mv^\mu\,\,.
\label
{p=mv-4}
\end{equation}                        

It follows from (\ref{p=mv-4}) that ${\dot p}^{\mu}=ma^\mu$.
In view of (\ref{four-velocity-times-four-acceleration=0}) and (\ref{f-cdot-v}),  the projector 
$\stackrel{\scriptstyle v}{\bot}$ in (\ref{newton4}) 
acts as a unit operator, and (\ref{newton4}) reduces to 
\begin{equation}
ma^\mu=f^\mu\,\,.
\label
{ma=f-4}
\end{equation}                        
This is the equation
of motion for a Galilean particle in four-dimensional vector notation.
Since 
$v\cdot a=0$ and $v\cdot f=0$, this equation contains  three independent components.

If $f^\mu=0$, then (\ref{ma=f-4}) becomes $\dot v^\mu=0$.
This equation has a unique solution $v^\mu={\rm const}$.
Thus, a free Galilean particle moves along {\it straight} timelike world lines.
This result can be used for formulating an alternative definition of Galilean
particles.

Let us translate (\ref{ma=f-4}) into the three-dimensional language.
In a particular inertial frame,
the four-momentum $p^\mu$ can be decomposed:
\begin{equation}
p^\mu=({\varepsilon},{\bf p}).
\label
{p-Galilei-decomp}
\end{equation}                        
When (\ref{four-velocity-in terms-velocity}) is compared with (\ref{p-Galilei-decomp}), 
it is apparent that
\begin{equation}
{\varepsilon}=m\gamma,
\label
{E-Galilei}
\end{equation}                        
\begin{equation}
{\bf p}=m\gamma{\bf v}.
\label
{p-Galilei}
\end{equation}                        
In the nonrelativistic limit ${\bf v}\to 0$, these expressions 
can be expanded in powers of ${\bf v}$:
\begin{equation}
{\varepsilon}=\frac{m}{\sqrt{1-{\bf v}^2}}=m+\frac{m{\bf v}^2}{2}+\ldots,
\label
{E-Galilei-n-rel}
\end{equation}                        
\begin{equation}
{\bf p}=\frac{m{\bf v}}{\sqrt{1-{\bf v}^2}}=m{\bf v}+\ldots.
\label
{p-Galilei-n-rel}
\end{equation}                        
The term $m{\bf v}$ on the right of (\ref{p-Galilei-n-rel}) is the 
Newtonian 
momentum, while the two terms 
on the right of (\ref{E-Galilei-n-rel})
differ from the conventional
nonrelativistic kinetic energy by $m$.
However, the energy in Newtonian mechanics is defined 
up to an arbitrary constant.
The first term on the right of (\ref{E-Galilei-n-rel})
is called the {\it rest energy}.
We see that the four-dimensional picture requires that the expansion of $\varepsilon$ 
begin with $m$. 
The quantity $m$ is not only a measure of
inertia 
of a Galilean particle, but also 
the total energy of 
this particle in its rest frame.
The famous relativistic relation 
\begin{equation}
\varepsilon_0={m}\,\,,
\label
{varepsilon=m}
\end{equation}   
expressing the `equivalence of mass and energy', is thus a mere consequence of the defining equation (\ref{p=mv-4}).
In the strict sense, equation (\ref{varepsilon=m}) only holds for Galilean particles.

Let us {introduce} the three-force ${\bf F}$
exerting on our particle in a fixed inertial frame of reference by writing the decomposition of $f^\mu$ in this frame:
\begin{equation}
f^\mu=\gamma\,({\bf F}\cdot{\bf v},\,{\bf F})\,\,.
\label{f-mu-decomp}\end{equation}                        
This definition of ${\bf F}$ takes into account the orthogonality of $f^\mu$ and  
$v^\mu$.
The Lorentz factor $\gamma$ is introduced in (\ref{f-mu-decomp}) for convenience sake.
If we recall (\ref{four-acceleration-in terms-velocity}), and express the proper time $s$ 
via laboratory time of this frame $t$  using (\ref{dt=gamma-ds}), then (\ref{ma=f-4}) becomes
\begin{equation}
\frac{d}{dt}\,(m\gamma)={\bf F}\cdot{\bf v}\,\,,
\label
{work}
\end{equation}                        
\begin{equation}
\frac{d}{dt}\,(m\gamma{\bf v})={\bf F}\,\,.
\label
{planck-eq}
\end{equation}                        

Equation (\ref {planck-eq}) describes 
 the variation of the 
momentum ${\bf p}=m\gamma{\bf v} $ of a relativistic Galilean particle affected by the force $ {\bf F} $, while equation (\ref {work}) is the relation between rate of change of energy 
$\varepsilon=m\gamma$ and rate of doing work ${\bf F}\cdot{\bf v}$.

Equation (\ref {planck-eq}) is often poorly referred to as the relativistic generalization of 
Newton's second law.
However, it is clear from the above that there is nothing `more general' in (\ref {planck-eq}) as 
compared to (\ref {newton-orig-p}).

\section{Conclusion and outlook (for the expert reader)}
\label
{concl}

The impossibility to create and annihilate  particles in the classical picture is a severely 
restrictive condition, whence it follows that the concept of a world line is very simple.
The set of allowable world lines contains only timelike and lightlike smooth curves.
Note that if this set of world lines represents histories of charged particles, this results in
a unique description of retarded electromagnetic fields.
A charged particle,  moving along a  
world line of this kind, generates a single-valued retarded field, 
the Li\'enard--Wiechert field.
These restrictions may not be weakened, otherwise the  retarded electromagnetic field is no longer single-valued.

These world lines are appropriate for 
formulating the least action 
principle with end points  separated by  timelike intervals.
The relevance of smooth timelike and null 
curves\footnote{Two points on a null smooth curve are, in general,
separated by a timelike interval (the exception is provided by  straight null curves).
As a simple example, we refer to the history of a particle which orbits 
in a circle of radius $r$ at an angular velocity
of ${\dot\varphi}=1/r$.
This history  is depicted by a helical null world 
line of radius $r$ wound around the time axis.
Any two points on the helix are separated by a timelike interval.
Null curves are discussed in greater detail in \cite{Bonnor}.} to the least action principle is 
 apparent, while  spacelike curves are unsuitable for this purpose.
Furthermore, we should abandon timelike $V$-shaped and $\Lambda$-shaped curves, shown in Figure \ref{crann}.  
Indeed,  a spacelike hyperplane (which is an instantaneous snapshot
of space in a particular inertial frame) may intersect a 
timelike $V$-shaped curve twice, otherwise it fails to intersect it 
at all. 
The same is true for $\Lambda$-shaped curves.
Therefore, the least
action principle with end points separated by timelike intervals cannot be unambiguously
formulated for such world lines. 

The dynamical law for a relativistic particle (\ref{newton4}) is Newton's second law 
projected on hyperplanes perpendicular to the world line.  
Therefore, in relativistic mechanics, Newton's second law requires neither modification nor generalization. 
It should be only smoothly embedded in the four-dimensional geometry of Minkowski spacetime.

Newton believed that equation (\ref{p=mv-3}) is 
the universal relation between the momentum 
and kinematical variables.
However, this is not the case.
The actual role of this equation is to define  
{Galilean particles}.
In the four-dimensional picture, the 
definition is reformulated: a {Galilean particle} is an 
object possessing the four-momentum $p^{\mu}$
 shown in (\ref{p=mv-4}). 

The dynamics of a Galilean particle is governed by equation (\ref{ma=f-4}).
The simplicity of this equation is especially striking if  $f^\mu$ is linear in  $v^\mu$  and independent of  spacetime, which is the case when a 
charged particle moves in a constant homogeneous electromagnetic field.
Then equation (\ref{ma=f-4}) is linear.
However, the simplicity is at once lost when this equation is expressed in three-dimensional vector notation, because (\ref{planck-eq}) is a nonlinear differential equation in all instances.

Two invariants can be built from $p^\mu$ and $v^\mu$,
\begin{equation}
M^2 =p^2\,\,
\label
{M-def}
\end{equation}                                          
and
\begin{equation}
{\rm m}=v\cdot p\,\,,
\label
{m-def}
\end{equation}                                          
while the third invariant $v^2=1$ is dynamically trivial. 
Appropriate names for 
$M$ and ${\rm m}$ are the {\it mass} and 
the {\it rest mass}, respectively.
For a Galilean particle these quantities are 
identical to each other and the Newtonian mass $m$.
Moreover, it follows from $v^2=1$ that $M$ is equal to $m$.
Thus, formulas (\ref{M-def}) and (\ref{m-def}) represent
two different definitions of the same quantity attributable to a 
Galilean particle.

However, the classical theory admits non-Galilean objects as well.
Such objects possess momenta $p^\mu$ which 
depend on kinematical variables 
differently than that given by (\ref{p=mv-4}), and 
the equations
of motion for those objects are not as simple in form as equation 
(\ref{ma=f-4}). 
In the case $f^\mu=0$, non-Galilean objects do not necessarily move along {straight} 
world lines but may execute vibrations around
a straight world line (the so-called zitterbewegung), or self-acceleration, or self-deceleration,
or complex combinations of these motions.
As an example of non-Galilean objects we refer to  a classical particle with spin, say, a tiny top whose size tends to zero \cite{Frenkel}.
The four-momentum of a free spinning particle in this model is given by 
\begin{equation}
p^\mu=\frac{M^2}{{\rm m}}\,v^\mu+\frac{{\bf S}^2}{{\rm m}}\,{\dot a}^\mu\,\,,
\label
{p-mu-spinning}
\end{equation}                                          
where the three-vector ${\bf S}$ is its spin as viewed by a comoving observer.
Since $p^\mu$ and $v^\mu$ are not 
parallel,  $M\ne {\rm m}$. 
For an extended discussion of  non-Galilean particles see \cite{k3}.


\begin{thebibliography}{99}
\label
{bibl}

\bibitem
{Minkowski}  
 Minkowski H (1908/9)  Raum und Zeit {\it Phys. Zeitschr.}  {\bf 10} 75-88  

\bibitem
{Abraham}
Pais A (1982)
{\it ``Subtle is the Lord...'' The Science and the Life of Albert Einstein} 
(Oxford: Oxford University Press) 
p 152
 
\bibitem
{Einstein13}
 Einstein A \&  Grossmann  M (1913) Entwurf einer verallgemeinerten Relativit\"atstheorie
und Theorie der Gravitation {\it Z. Math. und Phys.} {\bf 62} 225-261  

\bibitem
{Feynman}
Feynman R P (1966) The development of the space-time view of quantum electrodynamics
{\it Science} {\bf 153} 699-708

\bibitem
{knots}
Seifert H \& Threlfall  W (1934) {\it Lehrbuch der Topologie} (Leipzig: Teubner) [English translation:   
{\it A Textbook of Topology} (New York: AMS Chelsea, 1980)] 
pp 2 and 315

\bibitem
{Hartle77}
 Hartle J B (2003) {\it Gravitaty. An Introduction to 
Einstein's General Relativity} (San Francisco:  Addison Wesley) p 77

\bibitem
{Hartle85}
 Hartle J B (2003) {\it Gravitaty. An Introduction to 
Einstein's General Relativity} (San Francisco:  Addison Wesley) p 85

\bibitem
{Bonnor}
 Bonnor W B (1969) Null curves in a Minkowski space-time {\it Tensor, N. S.} {\bf 20} 229-242 

\bibitem
{Frenkel}
 Frenkel J (1926) Die Elektrodynamik des rotierenden Electrons 
{\it Z. Phys.} {\bf 37} 243-253 


\bibitem
{k3}
Kosyakov B P (2003) On the inert properties of particles in classical theory 
{\it Phys. Part. Nucl.} {\bf 34} 808-828 [Translated from {\it Fizika  Elementarnykh Chastits i Atomnogo Yadra} {\bf 34} 1564-1609] 




\end{thebibliography}
\end{document}